\documentclass[prb,showpacs,showkeys]{revtex4}
\usepackage{amsfonts}
\usepackage{amsmath}
\usepackage{amssymb}
\usepackage{graphicx}
\usepackage{subfigure}

\begin{document}

\title{Stability mechanism of cuboctahedral clusters in UO$_{2+x}$: First-principles calculations}
\author{Hua Y. Geng}
\affiliation{Department of Quantum Engineering and Systems Science, The University of Tokyo, Hongo 7-3-1, Tokyo 113-8656,
Japan}
\author{Ying Chen}
\affiliation{Department of Quantum Engineering and Systems Science, The University of Tokyo, Hongo 7-3-1, Tokyo 113-8656,
Japan}
\author{Yasunori Kaneta}
\affiliation{Department of Quantum Engineering and Systems Science, The University of Tokyo, Hongo 7-3-1, Tokyo 113-8656,
Japan}
\author{Motoyasu Kinoshita}
\affiliation{Nuclear Technology Research Laboratory, Central Research Institute of Electric Power Industry, Tokyo 201-8511, Japan}
\affiliation{Japan Atomic Energy Agency, Ibaraki 319-1195, Japan}

\begin{abstract}
The stability mechanism of cuboctahedral clusters in nonstoichiometric uranium
dioxide is investigated by first-principles LSDA+U method. Calculations reveal that
the structural stability is inherited from U$_{6}$O$_{12}$ molecular cluster
whereas the energy gain through occupying its center with an additional oxygen
makes the cluster win out by competition with point oxygen interstitials. Local displacement
of the center oxygen along $\langle111\rangle$ direction also leads the cluster
8-folded degeneracy and increases relatively the concentration at finite temperatures.
But totally, elevation of temperature, \emph{i.e.}, the effect of entropy, favors point interstitial over
cuboctahedral clusters.
\end{abstract}

\volumeyear{year}
\volumenumber{number}
\issuenumber{number}
\eid{identifier}
\maketitle


Uranium dioxide adopts the simple fluorite (CaF$_{2}$) type of crystal structure
with a space group $Fm\overline{3}m$. But its self-defects, as in most anion excess
fluorites, exhibit rather complex behavior: experimentally oxygen interstitials do not
occupy the largest cation octahedral hole ($\frac{1}{2}$,\,$\frac{1}{2}$,\,$\frac{1}{2}$) but
form low symmetric clusters composed of oxygen vacancies instead.\cite{willis64a, willis64b, willis78} The exact geometry
of these clusters, however, is unclear. Experimentalists have proposed several
structural models to explain the measured neutron diffraction patterns,\cite{willis78, murray90} of which the cuboctahedral
cluster (COT) appearing in U$_{4}$O$_{9}$ \cite{bevan86,cooper04} and U$_{3}$O$_{7}$ \cite{garrido03,nowicki00} is the most clearly described.
But even with this structure, ambiguity remains about whether the center is occupied by an additional oxygen (COT-o) or not (COT-v),\cite{garrido03} and
if the center atom really displaced off-center along $\langle111\rangle$ direction in the case that it was occupied.\cite{bevan86,cooper04}
Theoretical analysis have done little help so far since most of work confined
to point oxygen interstitial (O$_{i}$) and vacancy (O$_{v}$) where the former
always sites at the octahedral hole and thus failed to explain the experimental phenomena.\cite{crocombette01,freyss05,geng08}
Inspired by its close relationship with Willis type
clusters, COT was suggested also should present in UO$_{2+x}$.\cite{murray90, geng08}
A fully understanding of its property and stability mechanism thus becomes
important not
only for a general description of fluorite-related clusters \cite{bevan86b} but also
for nuclear applications, for example the safety disposal of used nuclear fuel where the formation of U$_{4}$O$_{9}$/U$_{3}$O$_{7}$
is a key process for the development of U$_{3}$O$_{8}$ phase which can lead to splitting of the sheath.\cite{mceachern98}

\begin{figure}
\centering
\begin{tabular}{|c|}
\hline
\subfigure[]{\label{fig:struct:a}\includegraphics[width=1.4 in]{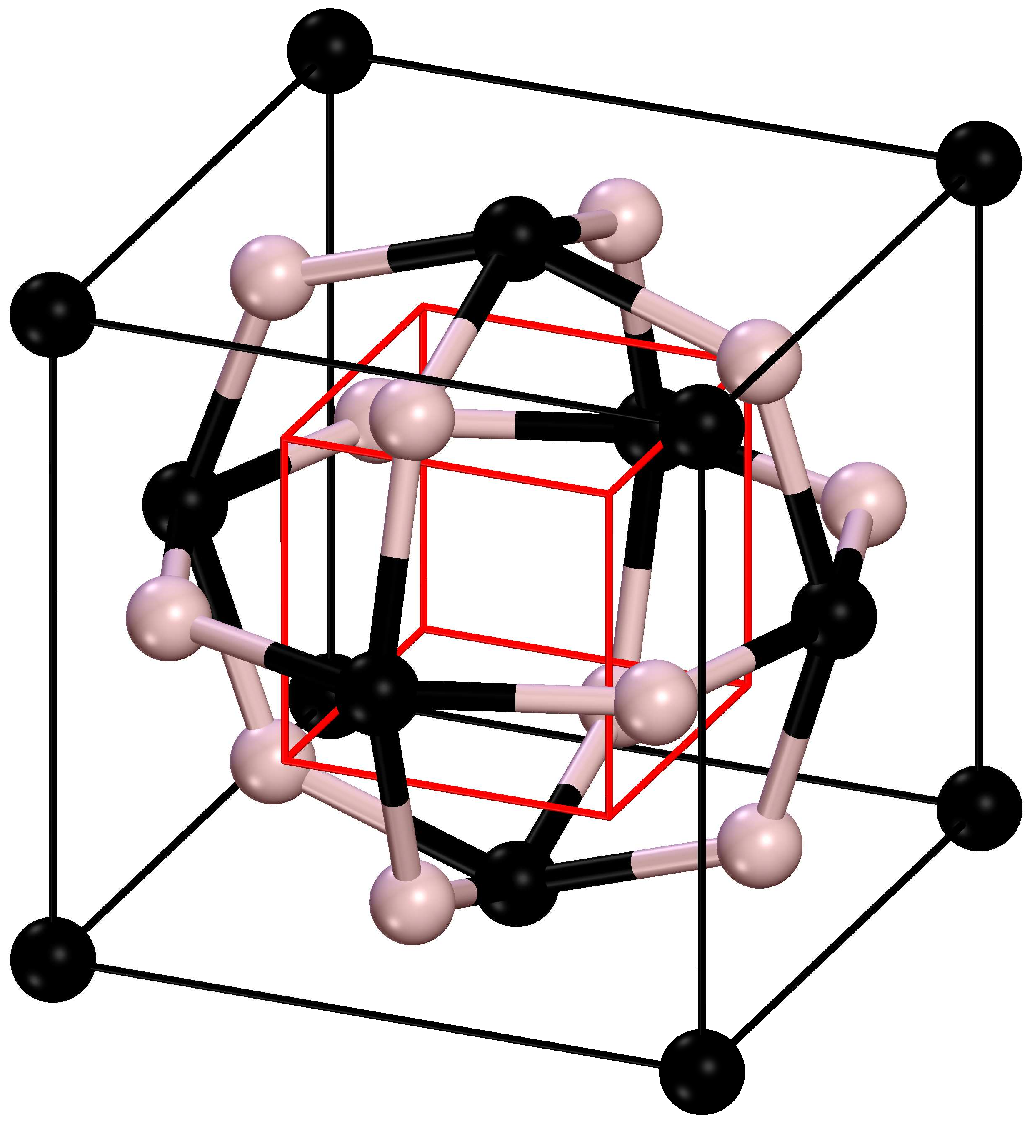}}
\hfil
\subfigure[]{\label{fig:struct:b}\includegraphics[width=1.4 in]{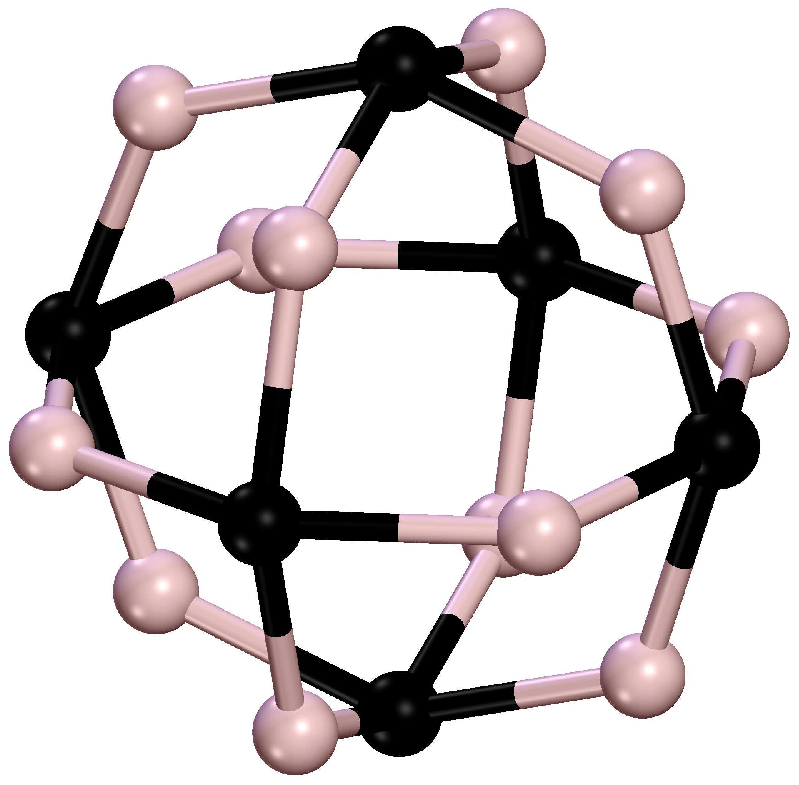}}\\
\hline
\subfigure[]{\label{fig:struct:c}{\raisebox{0.08 in}{\includegraphics*[width=1.25 in]{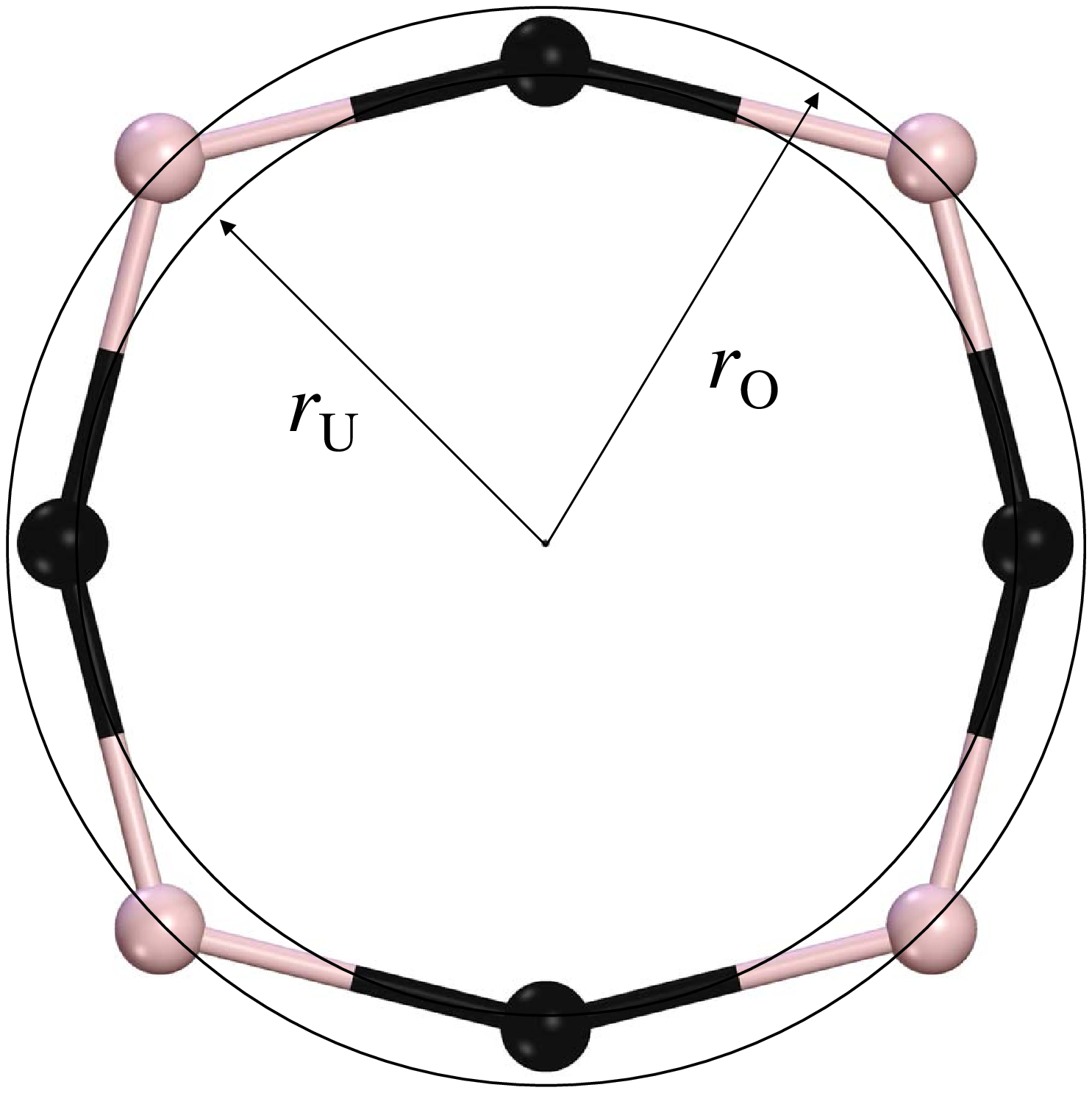}}}}
\subfigure[]{\label{fig:struct:d}\includegraphics[width=1.75 in]{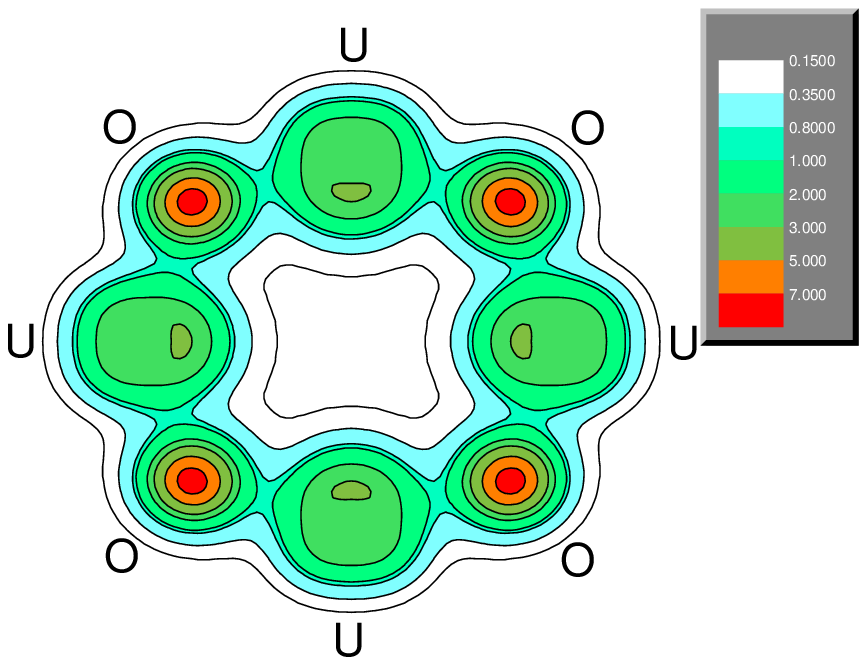}}\\
\hline
\end{tabular}
  \caption{(Color online). (a) Cuboctahedral cluster (COT-v) incorporated in a fluorite cell, where
  the small inner cube indicates the (removed) oxygen cage; (b)
  U$_{6}$O$_{12}$ molecular cluster; (c) One of the three mutually perpendicular U-O rings in
  U$_{6}$O$_{12}$ cluster,
  and (d) its charge density.}
  \label{fig:struct}
\end{figure}

Historically, COT was denoted by M$_{6}$X$_{36}$ (or M$_{6}$X$_{37}$ if an
additional anion occupies the center).\cite{bevan86,bevan86b} It is a little misleading since the actual
defect composed of only 12 interstitials [forming the cuboctahedral geometry, see
Fig.\ref{fig:struct:a}] and 8 vacancies [forming the small cube in Fig.\ref{fig:struct:a}]. Different from Willis type
clusters whose stability can be interpreted in a similar concept of split-interstitial
defect where several atoms share the common lattice sites,\cite{geng08}
COT is of more regular and with higher symmetry [point
group $O_{h}(m\overline{3}m)$], and poises as
the special one. In fact, our calculations that will be reported here revealed it is actually an U$_{6}$O$_{12}$
\emph{molecular cluster} incorporated in bulk fluorite UO$_{2}$ by sharing the uranium atoms
with the cation sublattice face centers [Fig.1(a-b)] after removed the eight corresponding lattice oxygens from the matrix.
The thus inserted 12 oxygens presented as Willis O$^{'}$ interstitials
that displaced along $\langle110\rangle$ directions in the picture of fluorite structure.

In calculations, COT cluster was modeled by embedding it into a cubic supercell of fluorite UO$_{2}$ with otherwise
96 atoms (U$_{32}$O$_{64}$). This configuration has large enough cell size
with a deviation composition $x=\frac{1}{8}$ for COT-v and $\frac{5}{32}$ for COT-o,
compared with experimental $x\simeq0.21$ for U$_{4}$O$_{9}$ \cite{cooper04} and $0.35$ for
U$_{3}$O$_{7}$,\cite{garrido03} respectively.
U$_{6}$O$_{12}$ molecular cluster was modeled by put into a vacuum cubic box with a lateral
length of 11\AA, a sufficient distance for current purpose. Total energies
were calculated with plane wave method based on density functional theory (DFT),\cite{kresse96}
with generalized gradient approximation (GGA, for U$_{6}$O$_{12}$), local
density approximation with Hubbard correction (LSDA+U,\cite{anisimov91,anisimov93} for COTs) and projector-augmented wave (PAW)
pseudopotentials.\cite{blochl94,kresse99} All structures have been fully relaxed to get all forces and stress less than 0.01\,eV/\AA.
Other computational parameters such as energy cutoff and sampling k-points are
the same as those in Ref.[\onlinecite{geng08}] which focused on point defects behaviors.

\emph{Geometry and energetics.---}The uraniums in COTs that embedded in UO$_{2}$ are found always
bond to interstitial oxygens firstly with a shorter bond length than to
the nearest neighbor (NN) lattice oxygens (2.2 \emph{vs} 2.4\,\AA).
Analysis of electronic density also shows the weak covalent bonds that forming an U$_{6}$O$_{12}$ cluster
are always prior to other bonds [Fig.\ref{fig:struct:d}]. Local distortions have not changed the picture
and preserve much of the tightly connected feature of vacuum U$_{6}$O$_{12}$
cluster, which
consists of three mutually perpendicular U-O rings that in turn determined by
the two radius from the cluster center to the uranium ($r_{U}$) and oxygen ($r_{O}$)
atoms completely, as shown in Fig.\ref{fig:struct:c}.
Vacuum U$_{6}$O$_{12}$ is perfect and without any deformation
freedom such as variation in U-O bond length and related angles.
All real but low vibrational frequencies (not shown here) indicate the structure is locally
stable but flexible for distortion. A cohesive energy of 21.2\,eV per UO$_{2}$ molecule
is comparable with the bulk material [22.3(23.9)\,eV of experiment(GGA)].\cite{geng07} These features
and the geometry make U$_{6}$O$_{12}$ can be incorporated naturally with fluorite
crystal (or generally, FCC lattice) and forms COT clusters without disturb the host structure severely.

\begin{table*}[]
\caption{\label{tab:Struct-E} First principles results for structural and
energetic properties of oxygen defects in uranium dioxide: $x$ is the deviation
from the stoichiometric composition, $\Delta V$ the
defect induced
volume change that averaged to per fluorite cubic cell;
\emph{r}$_{\mathrm{U}}$ and \emph{r}$_{\mathrm{O}}$ the structural parameters of cuboctahedral
cluster as indicated in Fig.\ref{fig:struct:c}; $L$, $\widehat{\mathrm{UOU}}$
and $\widehat{\mathrm{OUO}}$ are the averaged nearest neighbor bond length between U and
O atoms, and the corresponding angles in cuboctahedral cluster, respectively; E$_{f}$, E$_{ef}$ and E$_{pf}$ are the
overall defect formation energy, formation energy per excess oxygen and per oxygen Frenkel pair, respectively.
}
\begin{ruledtabular}
\begin{tabular}{l c c c c c c c c c c} 
   & $x$ & $\Delta V$(\AA$^{3}$) & \emph{r}$_{\mathrm{U}}$(\AA) & \emph{r}$_{\mathrm{O}}$(\AA) & $L$(\AA)&$\widehat{\mathrm{UOU}}(^{\circ})$&$\widehat{\mathrm{OUO}}(^{\circ})$& E$_{f}$(eV) & E$_{ef}$(eV) & E$_{pf}$(eV) \\
\hline
  U$_{6}$O$_{12}$  & 0   &       & 2.54 & 2.87& 2.09 & 118.3 & 151.7 &&&  \\
  COT-v  &$\frac{1}{8}$  & $-0.14$& 2.92 & 2.82& 2.20 & 139.7 & 129.8 &$-7.18$&$-1.80$&1.91\\
  COT-o  &$\frac{5}{32}$ & $-1.61$& 2.76 & 2.90& 2.18 & 127.9 & 141.6 &$-12.41$&$-2.48$ &1.94 \\
  COT\footnotemark[1] & 0.21   & $-1.95$  &2.79 &2.93& 2.20 & 127.9 & 141.5 &&&  \\
  O$_{i}$\footnotemark[2]  & $\frac{1}{32}$   & $-0.29$ &  & &&&&$-2.17$&$-2.17$&5.36 \\
  O$_{v}$\footnotemark[2] & $-\frac{1}{32}$    & 0.20 &  &  &&&&7.53&&5.36 \\
\end{tabular}
\footnotetext[1]{\,Experiment of $\beta$-U$_{4}$O$_{9}$ at 503\,K reported in Ref.[\onlinecite{cooper04}]}
\footnotetext[2]{\,Isolated point defects, Ref.[\onlinecite{geng08}]}
\end{ruledtabular}
\end{table*}

Locally, COTs repulse the NN lattice oxygens outwards slightly, but no evident distortion
on cations was observed. As listed in table \ref{tab:Struct-E}, the overall
volume is contracted but the cube that contains the COT is expanded greatly, with a
lateral length of $2r_{\mathrm{U}}$. Also, embedding U$_{6}$O$_{12}$ into UO$_{2}$ not only
swells the cube (with an increased $r_{\mathrm{U}}$ and $r_{\mathrm{O}}$) but also
leads to other local distortions. Usually the distorted U-O ring is not on the same plane any longer,
and has additional freedoms in U-O bond length (\emph{L}) and related angles ($\widehat{\mathrm{UOU}}$ and $\widehat{\mathrm{OUO}}$).
Their averaged values are listed in table \ref{tab:Struct-E} by compared with
vacuum U$_{6}$O$_{12}$ and experimental estimates of COT measured at 503\,K on $\beta$-U$_{4}$O$_{9}$.\cite{cooper04}
Obviously, COT-o agrees with the experimental one much better than COT-v in geometry.
With an additional oxygen occupied the center, COT-o decreases the values of $r_{\mathrm{U}}$ and $\widehat{\mathrm{UOU}}$
while lifts $r_{\mathrm{O}}$ and $\widehat{\mathrm{OUO}}$ significantly with respect to COT-v, identifying
it as the one that appeared in experiments.

In contrast to the symmetry anticipation,\cite{bevan86} the center oxygen in COT-o does not site at the
\emph{real} center. It displaces along $\langle111\rangle$ direction
of fluorite structure with a distance about 0.43\,\AA, compared well with Cooper and Willis'
estimate of 0.64\,\AA,\cite{cooper04} and stands as an O$^{''}$ interstitial. In vacuum U$_{6}$O$_{12}$ cluster,
such kind of off-center displacement is forbidden. The small $r_{\mathrm{U}}$ ensures
the energy minimum always at the cluster center. But the $r_{\mathrm{U}}$ of COTs
are enlarged by bulk UO$_{2}$ matrix which in turn shifts the energy minimum off center. By
displacing the center oxygen along $\langle111\rangle$ direction, the three NN uraniums within
the corresponding section of the COT shell are
pulled inwards {$\sim0.2$\,\AA} (while the three oxygens are pushed out slightly $\sim0.1$\,\AA) and reduced
the U-O bond length from 2.76\, to 2.44\,\AA. The $O_{h}(m\overline{3}m)$ symmetry is also broken
to $C_{3v}(3m)$. In contrast, the symmetry broken in COT-v is mainly due to
Jahn-Teller distortion, where one uranium atom out relaxed but another five shift
inwards, results in a $C_{4v}(4mm)$ symmetry.

The large cohesive energy of U$_{6}$O$_{12}$ leads to a deep formation energy
of COT clusters, as indicated in table \ref{tab:Struct-E}, where the energetic information
of isolated point defects (O$_{i}$ and O$_{v}$) are also given for comparison.
By compensating excess oxygens with
point vacancies, we find the formation energy per Frenkel pair in COT is just one third of
the isolated case. However, the energy gain for each excess oxygen exhibits different
behavior for COT-v and COT-o. With the contribution of the center oxygen, the latter has
a lower E$_{ef}$ than the point interstitial but the former is weighed down by the
electronic density cavity presented in the cluster center which costs the energy significantly.

\begin{figure}
  \includegraphics*[0.22in,0.19in][4.02in,2.92in]{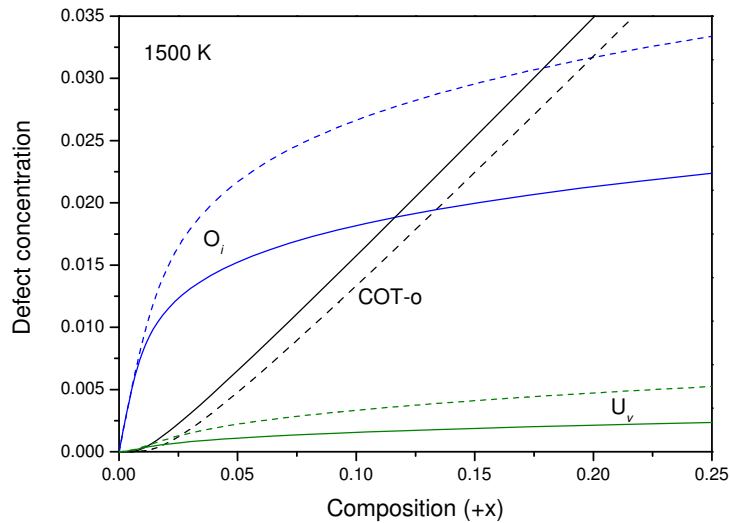}
  \caption{(Color online). Defect concentrations of point oxygen interstitial, uranium vacancy
  and COT-o cluster, respectively. Dashed lines indicate the corresponding results
  where the center oxygen
  in COT-o has no off-center displacement and thus with $g=1$.}
  \label{fig:deft-conc1}
\end{figure}

\emph{Defect concentrations.---}With regard to the concentration of COT at finite
temperatures, an intuitive picture is that it should favor moderate temperatures since
otherwise there have not sufficient vacancies to facilitate the formation
of the cluster. However, as mentioned above, the vacancies in COT
actually are not from point defects but inherited integrally from the U$_{6}$O$_{12}$ molecular cluster,
this naive picture thus becomes invalid. To calculate the defect concentrations at finite
temperatures properly, we employed here the independent clusters approximation (ICA)
(a generalization of the point defect model)\cite{geng08,matzke87,lidiard66}
in which all involved clusters are assumed to be
thermodynamically independent and obey Boltzmann distribution.
In closed regime where no particle-exchange with the exterior occurs,
the concentration $\rho_{i}$ of cluster $i$ that has an internal degeneracy $g_{i}$, $n_{i}$ excess oxygens
and a formation energy $E_{f}^{i}$ is given by
\begin{equation}
\rho_{i}[V_{O}]^{n_{i}}=g_{i}\exp\left(\frac{-E_{f}^{i}-n_{i}\times E_{f}^{O_{v}}}{\kappa_{B} T}\right).
\label{eq:ica}
\end{equation}
Here $n_{i}$ point oxygen vacancies (O$_{v}$) have been introduced as compensations.
Therefore we have a system contains point oxygen(uranium) defects (treated
as intrinsic Frenkel pairs), COT-v and COT-o
clusters that compete to each other. The oxygen and uranium subsystems are coupled up via
the isolated point Schottky defects.\cite{geng08} Among these defects, only COT-o has
an internal freedom with 8-folded degeneracy ($g=8$) arising from the $\langle111\rangle$ direction displacement of the center oxygen
and all others have $g=1$. COT-v has 4 and COT-o has 5 excess oxygen, respectively.
At a composition $x$ deviated from the stoichiometry, Eqs. (\ref{eq:ica})
are under a constraint of
\begin{equation}
  x=2([V_{U}]-[I_{U}])+[I_{O}]+\sum_{i}n_{i}\rho_{i}-2[V_{O}],
  \label{eq:conc}
\end{equation}
where the quantities in brackets denote point defect concentrations and $i$ runs over COT-v and COT-o clusters, with a coefficient equals to the
corresponding $n_{i}$ because each octahedral hole of the cation sublattice defines not only
a point interstitial site, but also a COT-v(o) cluster.

Using the calculated first-principles formation energies, we get the defect
concentrations as a function of temperature and composition
by solving Eqs. (\ref{eq:ica}) and (\ref{eq:conc}). The hypostoichiometric
regime is always dominated by O$_{v}$ and thus trivial. Interesting
competition between defects appears on the other section of composition with $x>0$.
Solid lines in Fig. \ref{fig:deft-conc1}
show the equilibrium concentrations of point oxygen interstitial O$_{i}$, uranium vacancy U$_{v}$ and
COT-o cluster at a temperature of 1500\,K around this regime. All other defects have a concentration
of ten more orders smaller and not shown here. O$_{i}$ is predominant
at low $x$ with
a rapid increment of its concentration, which flattens out gradually at high
compositions.
By contrast COT-o approaches a linear dependence on composition after $x>0.025$
and dominates the regime of $x>0.1$. The overall concentration of U$_{v}$ is
one order smaller than O$_{i}$ and has dismissed its influence on material properties.

The predominance of COT-o over COT-v is due to the energy gain of the center
oxygen but not the
entropy contribution of the 8-folded state. The dashed lines in Fig. \ref{fig:deft-conc1}
give the corresponding concentrations by treated COT-o with $g=1$. Removal
of the internal entropy contribution do reduce the competitiveness of COT-o
and increase the concentration of O$_{i}$ greatly, as well as that of U$_{v}$,
but has not changed the picture qualitatively---the concentration of COT-v is still
tens more orders smaller. Conversely, if rescale the formation energy of COT-o so that
it has the same E$_{ef}$ as COT-v, then they will have comparable concentrations,
but of ten orders smaller than that of O$_{i}$.

\begin{figure}
  \includegraphics*[0.22in,0.19in][4.08in,3.02in]{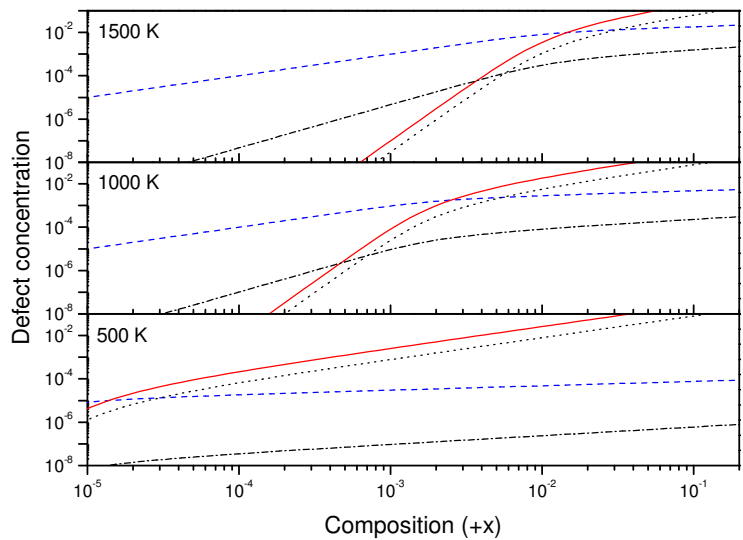}
  \caption{(Color online). Defect concentrations at different temperatures:
  solid(dotted) line---oxygen interstitial(vacancy) arising from COTs,
  dashed line---point oxygen interstitial, and dash-dotted line---point
  uranium vacancy. All other components are negligible.}
  \label{fig:deft-conc2}
\end{figure}

Figure \ref{fig:deft-conc2} shows the variation of defects competition
along temperature from 500 to 1500\,K. The solid(dotted) line indicates
the oxygen interstitial(vacancy) concentration arising from COT clusters:
each COT-o(v) contributes 13(12) oxygen interstitials and 8 vacancies.
It is obvious that increase
temperature, \emph{i.e.}, the entropy effect, favors point defects over COT clusters greatly.
The predominant range of O$_{i}$ has increased 3 orders by elevate the temperature
from 500 to 1500\,K. All point defect concentrations are enhanced along this process
except those from COT clusters, which are reduced by temperature.
This is because the probability to form a COT-o cluster from point defects
is proportion to $[I_{O}]^{13}[V_{O}]^{8}$ but each COT-o in conversely contributes only 13(8)
interstitials(vacancies), showing point defect is more disordered and with larger
entropy.

\emph{Summary and Discussion.---}Figure \ref{fig:deft-conc2} demonstrates
that at 500\,K COT-o is the exclusive defect cluster, support the
empirical assumption that $\beta$-U$_{4}$O$_{9}$ contains only this kind of cluster.\cite{cooper04}
From the temperature dependence of defect concentrations, we can be sure that
the lower temperature $\alpha$ phase also should contain COT-o exclusively.
On the other hand, the current interpretation of the neutron diffraction pattern in U$_{3}$O$_{7}$
is questionable, which employed COT-v instead of COT-o cluster.\cite{garrido03} The former
has a negligible concentration of ten more orders smaller and thus invalidates the analysis definitely.
The ordering
of the clusters that distributed in these phases,\cite{nowicki00} however, seems should be driven
by long-ranged strain energy rather than by chemical interactions. By the volume
change induced by defects listed in Table \ref{tab:Struct-E} and the fact that
COT itself expands the occupied fluorite cube seriously, there is a strong
deformation field around each COT cluster which repulses other COTs away.
The stress magnitude can be estimated from the bulk modulus of UO$_{2}$ as
$\sim2$\,GPa, a high enough value and any off-balance happened on the boundaries
of deformed domains will lead to cracks.
That is why U$_{4}$O$_{9}$/U$_{3}$O$_{7}$ film cannot protect UO$_{2}$ pellet
from being oxidized effectively.\cite{mceachern98} Such cracks are also believed as the onset
of high burn-up structures in nuclear fuels where uraniums are highly consumed and deteriorates the fuel quality
severely.\cite{noirot08}

\begin{acknowledgments}
Support from the Budget for
Nuclear Research of the Ministry of Education, Culture, Sports,
Science and Technology of Japan, based on the screening and counseling by the
Atomic Energy Commission is acknowledged.
\end{acknowledgments}


\end{document}